\documentclass[twocolumn,amsmath,amssymb,nofootinbib,superscriptaddress,longbibliography]{revtex4-1}

\usepackage{graphicx}
\usepackage{amsmath}
\usepackage{amssymb}
\usepackage{xcolor}
\usepackage{hyperref}
\usepackage{gensymb}
\usepackage{siunitx}
\usepackage{tabularx}
\hypersetup{pdfstartview={FitH},pdfpagemode={UseNone}, breaklinks=true, colorlinks=true, linkcolor=blue, citecolor=blue, urlcolor=blue, bookmarksopen=true, pdfnewwindow=true}
\usepackage[all]{hypcap}
\usepackage[english]{babel}
\usepackage{physics}

\makeatletter
\renewcommand{\fnum@figure}{FIG. \thefigure}
\makeatother

\begin{document}

\preprint{AIP/123-QED}

\title{Application of topological edge states in magnetic resonance imaging}

\author{Viktor M. Puchnin}

\author{Olga V. Matvievskaya}

\author{Alexey P. Slobozhanyuk}

\author{Alena V. Shchelokova}%

\altaffiliation{These authors contributed equally to this work}

\author{Nikita A. Olekhno}
\altaffiliation{These authors contributed equally to this work}
\affiliation{School of Physics and Engineering, ITMO University, Saint Petersburg 197101, Russian Federation}

\date{\today}

\begin{abstract}

Topological edge states in electromagnetic systems feature a set of attracting fundamental properties and unveil prospective applications based on disorder robustness and tailored localization. Despite active efforts in implementing topologically-protected waveguides in 2D photonic systems, applications of 1D topological systems remain almost uncharted. This letter demonstrates that topological edge modes can be realized in metamaterial-inspired volumetric resonators with a practical application in clinical magnetic resonance imaging (MRI). Performing numerical simulations and experiments with a 1.5 T MR scanner, we reconstruct the associated topological edge mode profiles and demonstrate their feasibility for sensitivity enhancement of conventional radiofrequency coils.
\end{abstract}

\maketitle
\def\thefootnote{}\footnotetext{Corresponding author: a.schelokova@metalab.ifmo.ru}


Topological photonics addresses edge- and corner-localized excitations whose existence is governed by general symmetries of a supporting structure rather than by any specific modifications of its boundary~\cite{2014_Lu}. Such symmetry protection leads to intriguing fundamental properties, including bulk-boundary correspondence, and induces increased robustness of topological edge states towards various imperfections in the host structure~\cite{2019_Ozawa}. There are numerous realizations of topologically-protected edge states in electromagnetic systems ranging from classical~\cite{2013_Hafezi, 2017_Bahari, 2019_Kruk} and quantum~\cite{2018_Blanco_Redondo_Science, 2018_Mittal_Nature} optics to gigahertz setups~\cite{2018_Peterson, 2018_Li, 2020_Li} with possible applications in signal multiplexers~\cite{2022_Nagulu} and even low-frequency electrical circuits~\cite{2015_Ningyuan, 2015_Albert, 2018_Rosenthal, 2019_Serra_Garcia_Circuit, 2020_Olekhno}. 

While the majority of the photonic topological insulators are designed to control edge~\cite{Wang2009} and surface~\cite{Yang2019} states, the abilities of 1D topological structures~\cite{Zhirihin2021} to sculpt electromagnetic near fields in real-life applications remain limited and are represented mostly by wireless power transfer~\cite{2021_Song, 2021_Zhang} and analog signal processing~\cite{2019_Zangeneh_Nejad, 2020_Zangeneh_Nejad}. Assessing the potential of topological edge states in 1D structures to strongly localize and manipulate the field at the subwavelength scale can make them an essential platform for practical applications, for example, to improve magnetic resonance imaging (MRI) characteristics through local enhancement of the transmit efficiency as well as the increase in the sensitivity of the radiofrequency (RF) coils~\cite{Webb2022}. In particular, it was shown that a metamaterial-based resonator could be used for the realization of a wireless coil for breast MRI~\cite{Shchelokova2020, 2021_Puchnin}. However, the previously demonstrated design has limitations with required near-field profiles preventing full coverage of fibroglandular tissue, which may lead to limitations for a very comprehensive assessment.


\begin{figure}[b]
    \centerline{\includegraphics[width=8.6cm]{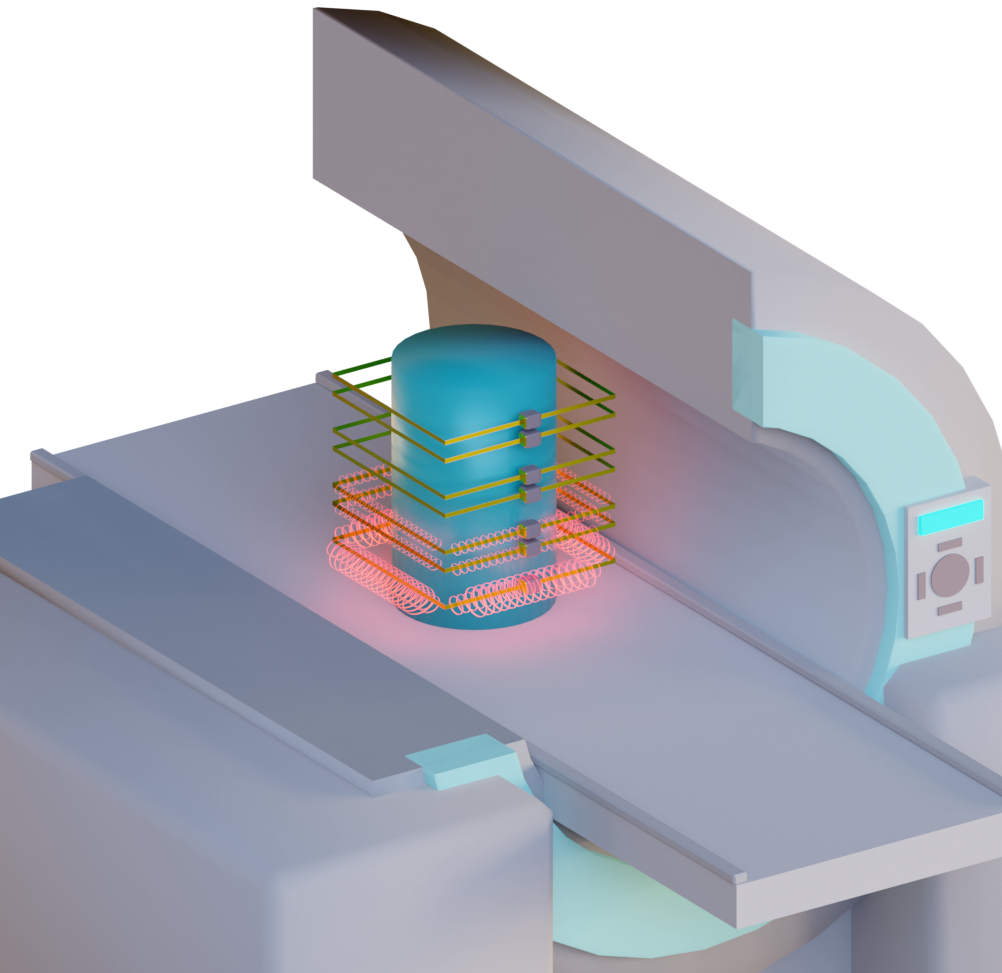}}
    \caption{Schematic view of a metamaterial-inspired volumetric resonator supporting a topological edge state (shown in red) together with a breast phantom (shown in teal color) located inside an MR scanner.}
    \label{fig:MRI}
\end{figure}


\begin{figure*}[tbp]
    \centering
    \includegraphics[width=16.5cm]{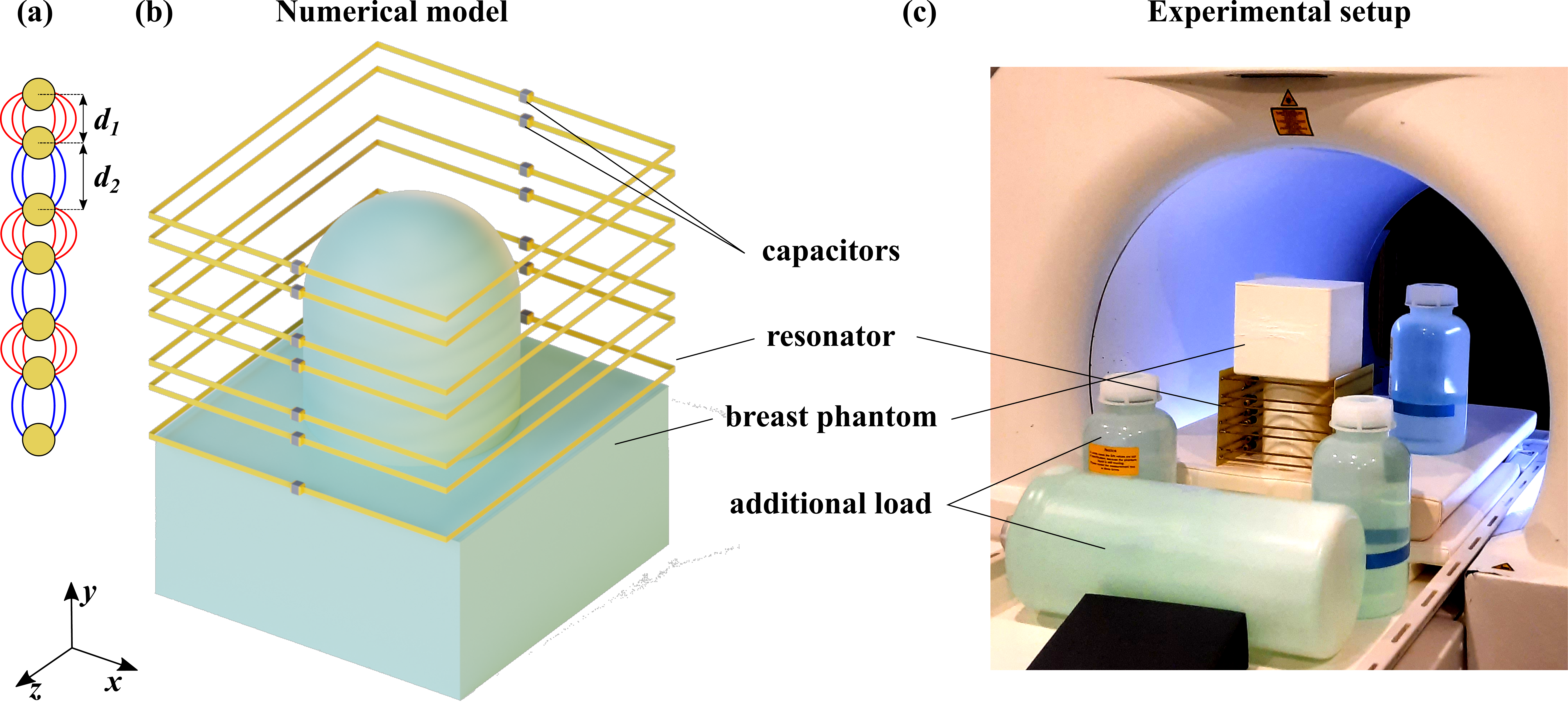}
    \caption{(a) Schematic view of the 1D Su-Schrieffer-Heeger model with elements placed at alternating distances $d_{1}$ and $d_{2}$. (b) Numerical model of the topological metasolenoid (yellow wires with capacitors) including a breast phantom (teal color). (c) Photograph of the experimental setup, featuring topological metasolenoid, the breast phantom, and additional load placed on the MR scanner table (different size bottles placed around the breast phantom).}
    \label{fig:Resonator}
\end{figure*}

In the present Letter, we demonstrate that topological edge states can be effectively implemented in a metamaterial-inspired volumetric resonator (dubbed \textit{metasolenoid})~\cite{2005_Maslovski} and used to overcome previous challenges in the resonator's design for MRI (Fig.~\ref{fig:MRI}). Such a device inductively couples with a quadrature RF whole-body birdcage coil of an MR scanner and focuses its magnetic flux in a desired small region, maximizing the transmit efficiency and receive sensitivity of the body coil. Inducing a geometrical dimerization in such arrays of strongly coupled split-ring resonators (SRRs), we implement Su-Schrieffer-Heeger (SSH) model~\cite{1979_SSH} and demonstrate that the topological edge mode emerges even in the presence of electromagnetic coupling between the considered metasolenoid and a body coil which excites circularly polarized RF magnetic field ($B_{1}^{+}$) in $1.5$~T MR scanners. Moreover, the topological resonator itself does not affect the polarization and magnitude of the RF magnetic field in the outer regions (i.e., does not disturb the functionality of the MR scanner), which allows probing such states with standard imaging pulse sequences and unveils prospects for their use in MRI. Finally, we estimate the signal-to-noise ratio (SNR) to demonstrate the efficiency and safety of the proposed topological metasolenoid.

We start with considering a plain (topologically trivial) metasolenoid similar to those studied in Refs.~\cite{2018_Shchelokova, 2021_Puchnin}, which consists of $7$ SRRs that are placed periodically with $d=18.5$~mm spacing forming an array with the height of $114$~mm. Each split-ring resonator is formed by a rectangular copper frame with dimensions of $156 \times 175~{\rm mm}^2$ and two gaps at the centers of the opposite sides. Then, a variable capacitor with a capacitance in the range $1-20$~pF is placed in each gap to adjust the resonator's resonance frequency to the Larmor frequency of the MR scanner. As demonstrated in Refs.~\cite{2018_Shchelokova, 2021_Puchnin}, such a resonator supports a highly homogeneous magnetic field in its volume, reaching a maximal amplitude at its center (see the supplementary material Figure S1 for details).

However, focusing RF fields at the metasolenoid edge (or even outside of it), as demonstrated in Fig.~\ref{fig:MRI}, might also be of interest for various applications (e.g., studying body region in the chest wall). While the standard metasolenoid design with evenly spaced SRRs clearly resembles Hubbard model~\cite{Pinto2009} with eigenmodes delocalized over the bulk, its dimerized analog with alternating distances $d_{1}$ and $d_{2}$ should implement the celebrated SSH model~\cite{1979_SSH} [Fig.~\ref{fig:Resonator}(a)]. Such a model, in turn, demonstrates the presence of a spectral bandgap hosting an in-gap state localized at the edge and topologically protected by the chiral symmetry~\cite{2019_Ozawa}.

Introducing a dimerization, we consider the metasolenoid with $n=7$ SRRs and the height of $114$~mm, keeping spatial dimensions of the structure the same as in Ref.~\cite{2021_Puchnin}, Fig.~\ref{fig:Resonator}(b,c). The distances between the nearest SRRs are $d_{1}=12.3$~mm and $d_{2}=24.7$~mm, respectively. In this case, the edge with the weak coupling $d_{2}$ features the presence of the topological edge state, in contrast to the edge with the strong coupling $d_{1}$~\cite{2019_Ozawa}. As shown in Fig.~\ref{fig:Resonator}(b), we consider the geometry with the weak coupling located at the breast-body transition to concentrate the fields under the breast.

Numerical simulations of the topological metasolenoid were performed in CST Studio Suite 2021 using the eigenmode and time domain solvers. A volumetric resonator was loaded with a homogeneous phantom simulating the properties and dimensions of the breast and the part of the body with the following properties: the size of the body area is $160 \times 160 \times 80~{\rm mm}^3$, the dielectric constant of the body is $\varepsilon=78$, the conductivity is $\sigma=0.45$~S/m; breast phantom parameters: the radius is $4.7$~cm, the height is $10.34$~cm, the dielectric constant is $\varepsilon=70$, and the conductivity is $\sigma=0.2$~S/m.

\begin{figure}[tbp]
    \centering
    \includegraphics[width=6cm]{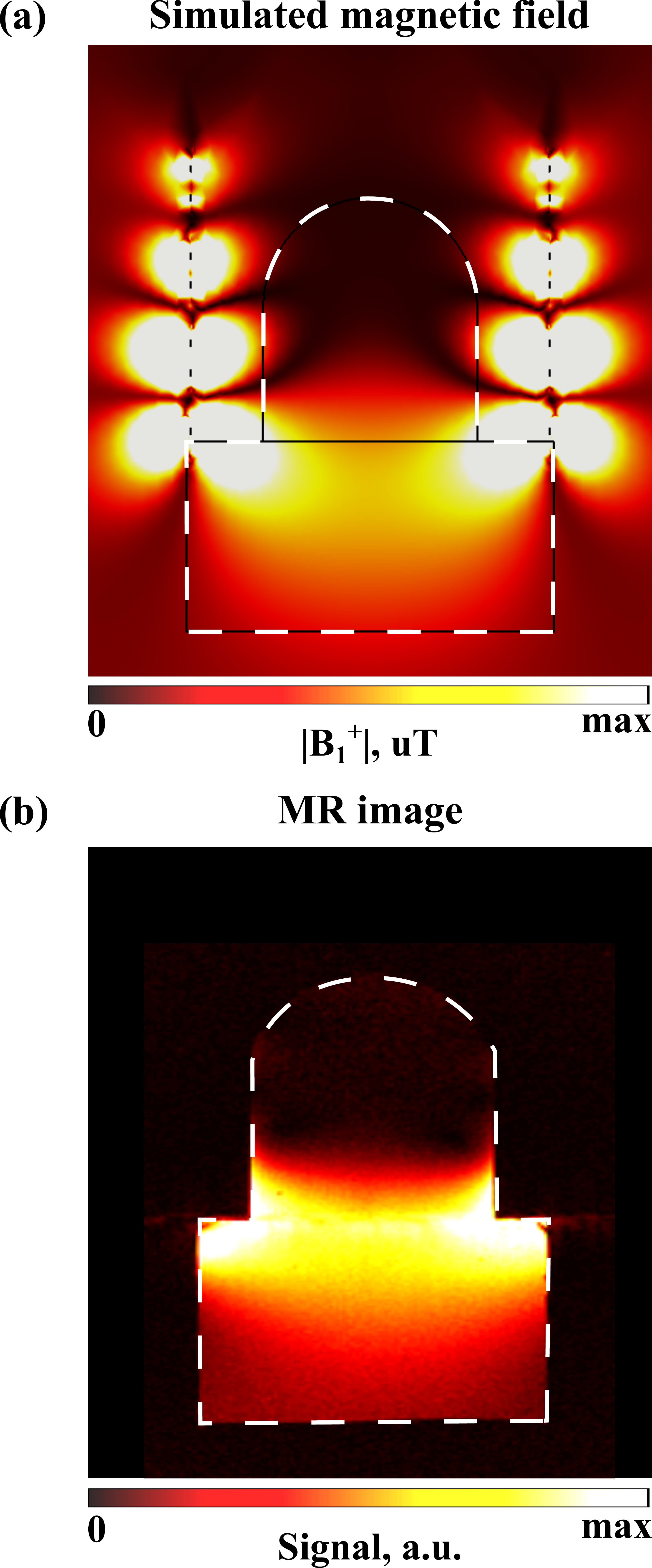}
    \caption{(a) Numerical simulation results of the edge-localized mode showing the amplitude of RF magnetic field $B_{1}^{+}$ at the frequency $f=63.68$~MHz. The topological resonator is excited by the circularly polarized magnetic field induced by the four waveguide ports. (b) Experimentally obtained MR image of the breast phantom.}
    \label{fig:Results}
\end{figure}

The standard metasolenoid has several eigenmodes (see the supplementary material Figure S1), which differ in current and electromagnetic field distributions~\cite{Jylh2005}. In addition, the uniform placement of SRRs is characterized by a symmetrical pattern of field distributions. The introduced dimerization qualitatively changes only the distribution of the fourth mode: the magnetic field concentrates near the edge SRRs, and in the rest of the resonator, the respective mode has an extremely low amplitude [Fig.~\ref{fig:Results}(a)] characteristic of the SSH chain.

Most clinical MR scanners use an RF birdcage body coil to excite the MR signal, creating a rotating transverse RF magnetic field within its volume (see the supplementary material Figure S2). As a part of the numerical simulation, we used an emulator consisting of four pairwise orthogonal waveguide ports that allow us to create a similar circularly-polarized magnetic field. Each port was given a phase delay of $90^{\circ}$ from the previous one to create a rotating magnetic field. The modified resonator loaded by a homogeneous phantom was placed at the center between all the ports. At the same time, the fourth resonant mode was tuned to $63.68$~MHz by selecting the proper capacitance of variable capacitors. Thus, we demonstrate that the topologically protected edge mode of the SSH model can be excited in a rotating RF magnetic field. Due to the inductive coupling of the topological metasolenoid with the exciting ports, the magnetic flux is focused inside the area of interest, i.e. outside the internal volume of the resonator, Fig.~\ref{fig:Results}(a). In addition, the maximum is located at a distance of $60$~mm from the edge SRR.

The experimental prototype of the topological metasolenoid includes two arrays of $14 \times 2$ copper strips (each with sizes $75 \times 3~{\rm mm}^{2}$) printed on two dielectric FR4 substrates with sizes of $164 \times 114 \times 1~{\rm mm}^{3}$ characterized by the permittivity $\varepsilon=4.3$ and losses $\tan\delta=0.003$ at the frequency of $63.68$~MHz. Each pair of copper strips on the same printed circuit board (PCB) is connected by variable capacitors with $C=4-20$~pF. The distances between pairs of SRRs are the same as used in numerical studies ($d_{1}=12.3$~mm and $d_{2}=24.7$~mm). Two PCBs are connected using telescopic brass tubes. The breast phantom shell is 3D-printed using polylactic acid plastic (PLA) and has a height of $100$~mm and a base radius of $50$~mm. The phantom solution consists of distilled water, NaCl (0.75 g/l), agarose (10 g/l), and a gadolinium-based MRI contrast agent (0.5 ml/l) to reduce spin-lattice relaxation time $T_1$. The phantom permittivity and conductivity were measured to be $\varepsilon = 78$ and $\sigma = 0.19$~S/m at $63.68$~MHz using a coaxial probe (DAK-12, SPEAG, Zurich, Switzerland).

For tuning the fourth mode to the Larmor frequency, we measured the reflection spectra of the topological metasolenoid using a non-resonant loop antenna connected with a vector network analyzer Planar (s5048, Chelyabinsk, Russian Federation). The loop antenna was placed at the center of the edge SRR. The fourth peak on the spectrum corresponds to a topologically protected resonant mode (see the supplementary material Figure S3). 

MRI experiments were performed on a $1.5$~T clinical MR scanner (Siemens MAGNETOM Espree) at Federal Almazov North-West Medical Research Centre (Saint Petersburg, Russian Federation). The breast phantom was used as the object of the study, the experimental setup is demonstrated in Fig.~\ref{fig:Resonator}(c). We also used a standard set of calibration body phantoms (Siemens) for additional load. The topological metasolenoid was placed on the patient table with several foam pads for its positioning at the center of the MR scanner bore. The axis of symmetry of the resonator was in the plane of rotation of the transverse RF magnetic field of the whole-body coil embedded into the scanner bore. MR scanner’s reference voltage ($U_{\rm ref}$) was calibrated to ensure that the actual flip angle in the region of interest (ROI) was equal to $180$ degrees. MR images of the breast phantom were obtained at the axial plane using a $T_{1}$-weighted gradient echo sequence with the following parameters: echo time (TE)=4.76~ms, repetition time (TR)=2500~ms, flip angle=20\degree, Field-Of-View=$260\times320~{\rm mm}^2$, matrix=$256\times208$. The resulting MR image is shown in Fig.~\ref{fig:Results}(b). The signal distribution on the image has the same behavior as in the numerical results: it localizes between the breast and body parts, demonstrating an excitation of the topologically protected edge state in an SSH array of SRRs using a clinical $1.5$~T MR scanner.

\begin{figure}[htbp]
    \centerline{\includegraphics[width=8.6cm]{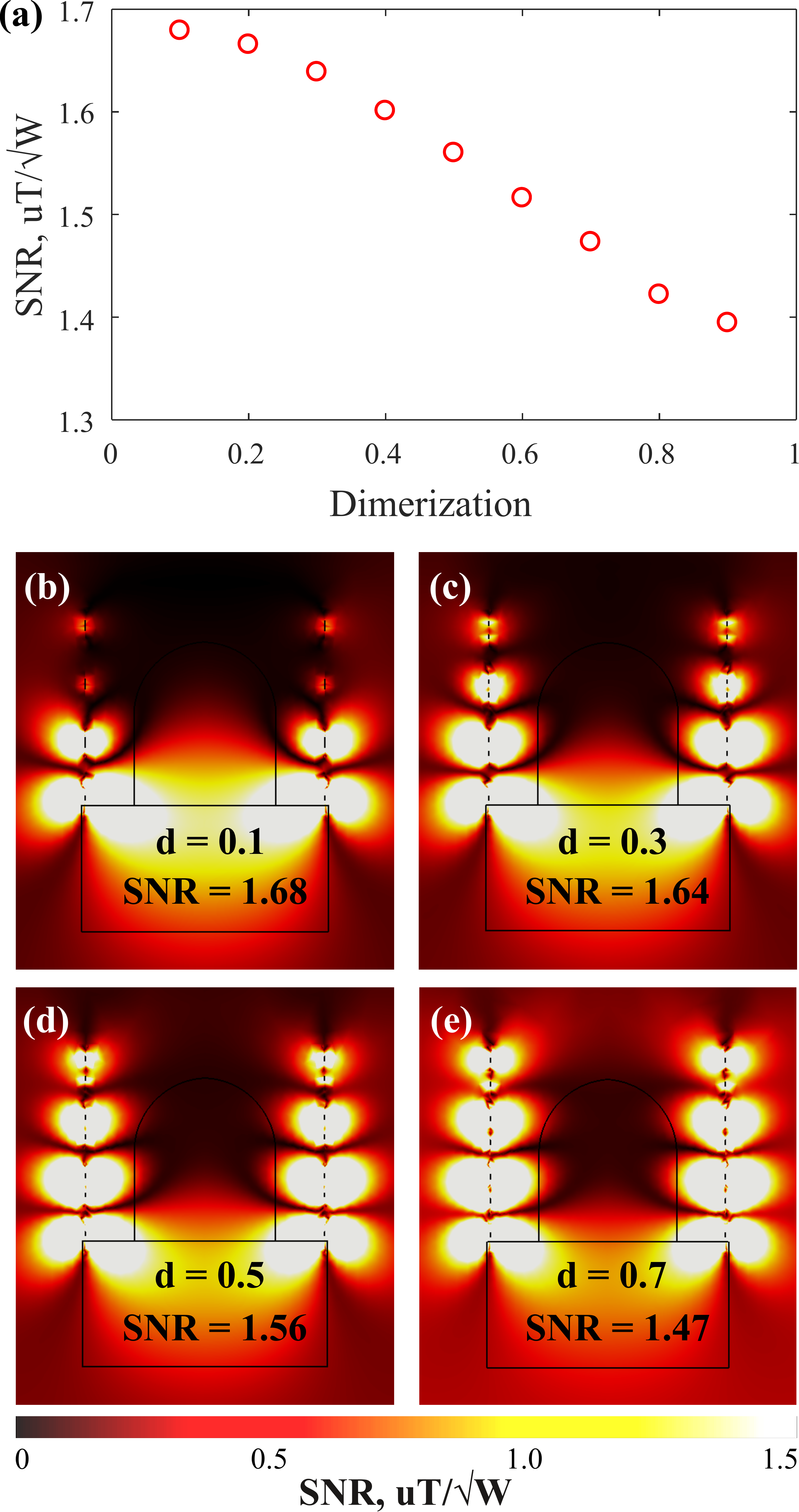}}
    \caption{(a) Signal-to-noise ratio (SNR) dependence on the dimerization degree. (b) Numerical model of the topological metasolenoid (yellow wires with capacitors), including a breast phantom. (b-e) SNR distributions for different dimerization degrees $d$: (b) $d=0.1$, (c) $d=0.3$, (d) $d=0.5$, and (e) $d=0.7$.}
    \label{fig:Optimization}
\end{figure}

In the next stage, we consider numerically the ways to increase the efficiency of the proposed structure in terms of the SNR in the body region. In high-field MRI systems, noise is proportional to the square root of the absorbed power, and the useful signal is proportional to the magnetic field. Therefore, in the first approximation, the SNR can be estimated as the ratio of the average value of the amplitude of the RF magnetic field to the square root of the power absorbed in the phantom. Figure~\ref{fig:Optimization}(a) presents the SNR dependence on dimerization. Different configurations were considered by changing the distances between SRRs as a ratio $d_{1}/d_{2}$ where $d_{1}$ is the distance inside SRRs pairs and $d_{2}$ is the distance between SRRs pairs. The mean value of SNR was calculated in the body part of the phantom. The mean SNR reduces inside the phantom body with a further increase in the homogeneity of SRR's structure. 

Figure~\ref{fig:Optimization}(b)-(e) demonstrates SNR maps in the central slice of the phantom for four different values of dimerization. It is seen that for higher contrast between $d_{1}$ and $d_{2}$, the SNR in the body phantom (imitating the chest wall area) becomes higher as well, facilitating an increase in the SNR for topological metasolenoid compared to the trivial one. These results are also supported by the measurements of the dependence of magnetic field amplitude on dimerization, showing an increase in the focusing of the magnetic field at the edge ring for higher contrasts between $d_{1}$ and $d_{2}$ (see the supplementary material Figure S4).


In this letter, we introduce a metamaterial-inspired resonator for MRI supporting a topologically protected edge mode and described by the 1D SHH model. Numerical simulations and experimental studies with a phantom reveal a strong localization of the RF magnetic field inside the body region under the breast, which is of key interest for searching breast cancer metastases~\cite{Gao2017}. We reveal an increase in SNR in the chest wall area for topological metasolenoid compared with the trivial one.

Prospective directions of further development include testing other quasi-one-dimensional topological models, including zig-zag arrays of SRRs~\cite{2014_Poddubny, 2019_Kruk, 2022_Kurganov}, engineering the effects of long-range couplings~\cite{2020_Li, 2021_Poshakinskiy, 2022_Olekhno_D4}, and considering non-linear effects in 1D SSH model~\cite{2018_Hadad} to open further degrees of tunability. Assessing the implementation of $\mathcal{PT}$-symmetric 1D SSH model with additional gain in SRRs~\cite{2020_Lazarides} and topological metasolenoids with external optical switching~\cite{2022_Kurganov} also looks very perspective.

\begin{acknowledgments}
The authors acknowledge valuable discussions with Dr. Dmitry Zhirihin. A.P.S. acknowledges the support from the Foundation for Advancement of Theoretical Physics and Mathematics ``BASIS.'' The work was supported by the Ministry of Science and Higher Education of the Russian Federation (project No. 075–15-2021–1391).
\end{acknowledgments}



\bibliography{Topology_BibFile}

\end{document}


\section*{Supplementary material}


\begin{center}
\Large\textbf{Application of topological edge states in magnetic resonance imaging}
\end{center}

\section{Comparison of eigenmodes of standard and topological metasolenoids}

\begin{figure*}[htbp]
    \renewcommand{\thefigure}{S\arabic{figure}}
    \centering
    \includegraphics[width=17 cm]{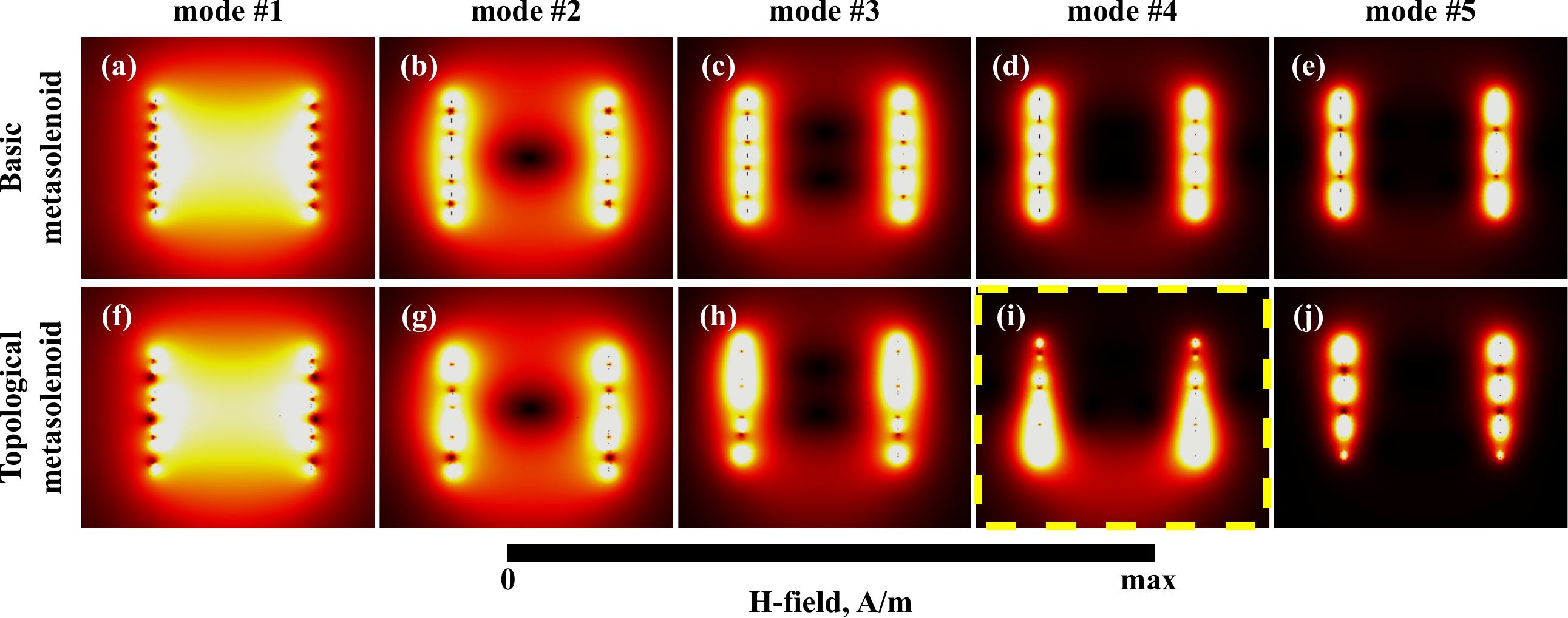}
    \caption{Numerically calculated $H$-field distributions of the different eigenmodes for the standard geometry of the metasolenoid (a-e) and for the topological metasolenoid (f-j). The yellow dashed line indicates the topologically protected edge state.}
    \label{fig:modes}
\end{figure*}

To compare the magnetic field distribution of the standard metasolenoid with $7$~SRRs and a topological metasolenoid (with dimerization in SRR spacings), we perform a numerical study using Eigenmode Solver in CST Microwave Studio 2021. Both metasolenoids have the same geometric dimensions and considered without any phantoms inside. We compare the magnetic field distributions for the first $5$ eigenmodes. The results presented in Fig.~\ref{fig:modes} show that modes \textnumero 1,2,3, and 5 have a weak difference in the magnetic field distribution, Fig.~\ref{fig:modes}. At the same time, mode \textnumero 4 for the topological metasolenoid has a strongly asymmetric magnetic field distribution localized near the edge ring characteristic of a topologically protected edge state, Fig.~\ref{fig:modes}(i).

\section{Excitation of a topologically protected edge state using a standard RF coil}

To investigate the possibility of excitation of a topologically protected edge state in the RF system of a clinical MR scanner, we perform numerical studies using a 16-leg shielded high-pass quadrature whole-body birdcage coil with the inner diameter of $70$~cm and the length of $65$~cm, tuned and matched to the frequency $63.68$~MHz with a homogeneous phantom placed in its center. A circularly polarized RF magnetic field ($B_{1}$) is created by two feeding ports with 90$^{\circ}$ phase shift. The phantom is the same as described in the main text (the size of the body area is $160\times160\times80~{\rm mm}^3$, the dielectric constant of the body is $\varepsilon=78$, the conductivity is $\sigma=0.45$~S/m; breast phantom size: the radius is $4.7$~cm, the height is $10.34$~cm, the dielectric constant is $\varepsilon=70$, and the conductivity is $\sigma=0.2$~S/m). 

\begin{figure*}[htbp]
    \renewcommand{\thefigure}{S\arabic{figure}}
    \centering
    \includegraphics[width=16.5cm]{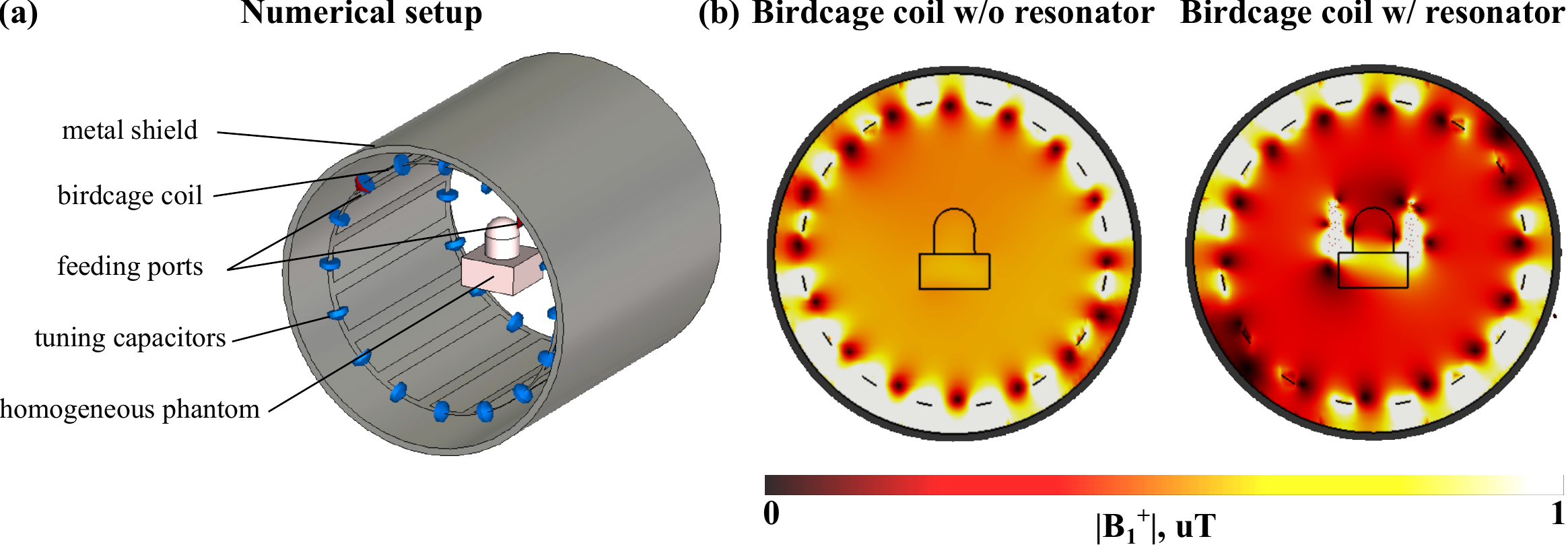}
    \caption{(a) View of the numerical model which includes a homogeneous phantom placed inside the whole-body birdcage coil model. (b) Numerically calculated $|B_{1}^{+}|_\mathrm{RMS}$ maps for $1$~W of total accepted power without (left) and with (right) the topological metasolenoid.}
    \label{fig:bc}
\end{figure*}

We compare two cases: with and without the topological metasolenoid placed around the breast phantom. The comparison is carried out by calculating the root-mean-square value (RMS) of the $\rm{B_1^+}$-field ($|\rm{B_1^+}|_{\rm{RMS}}$) per $1$~W of the total accepted power for each case in the breast area. The results demonstrate that using a standard body RF coil, it is possible to excite a topologically protected edge state in the metasolenoid. Such a resonator focuses a transverse RF magnetic field $B_1^+$ at the border of the phantom's body and breast Fig.~\ref{fig:bc}.

\section{The reflection spectrum of a topological metasolenoid.}

\begin{figure*}[htbp]
    \renewcommand{\thefigure}{S\arabic{figure}}
    \centering
    \includegraphics[width=8.6cm]{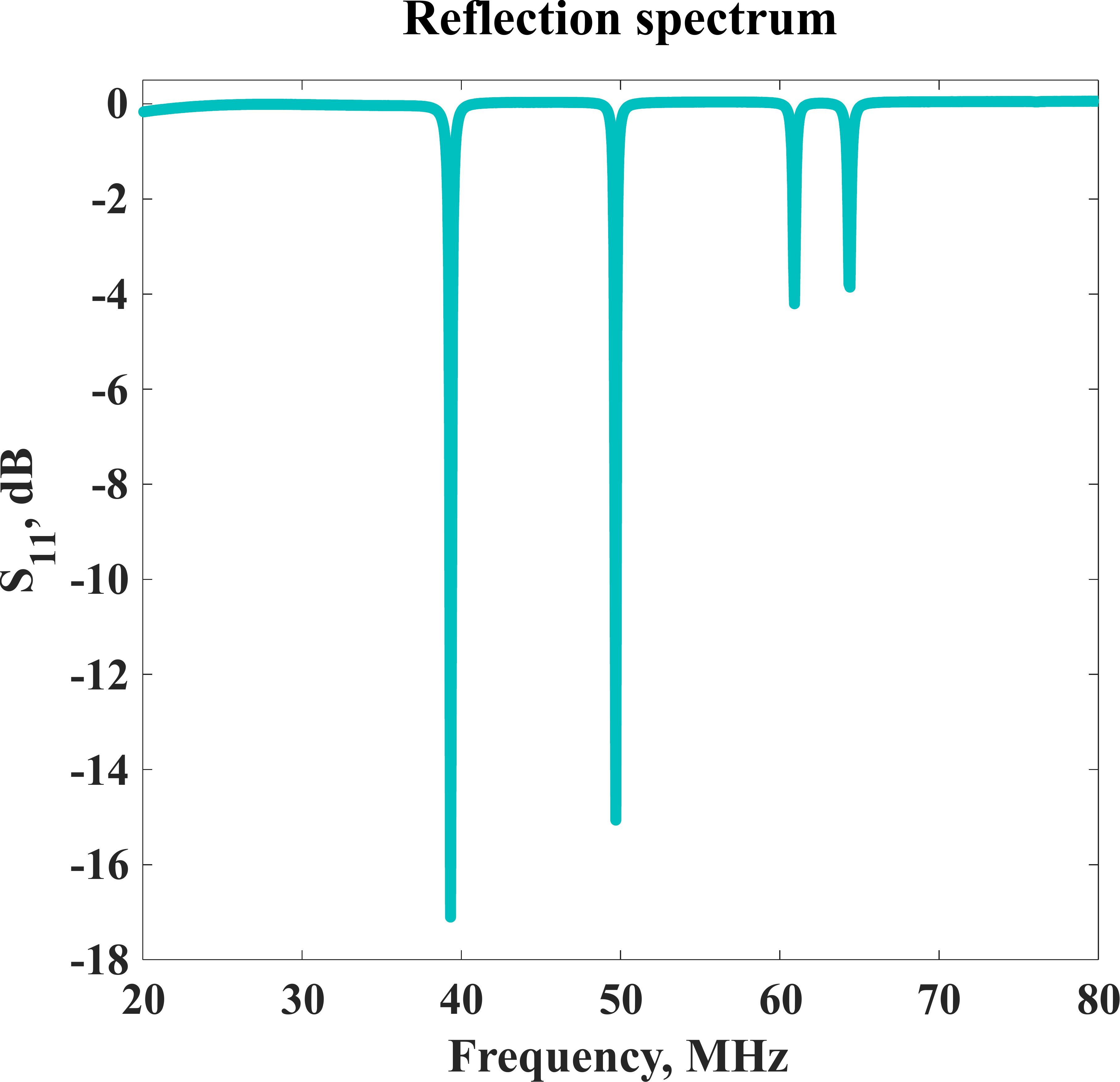}
    \caption{Experimentally measured reflection spectrum of the topological metasolenoid. The fourth peak corresponds to a topologically protected edge state.}
    \label{fig:S11}
\end{figure*}

\section{Dependence of the magnetic field amplitude on dimerization}

\begin{figure*}[htbp]
    \renewcommand{\thefigure}{S\arabic{figure}}
    \centering
    \includegraphics[width=9cm]{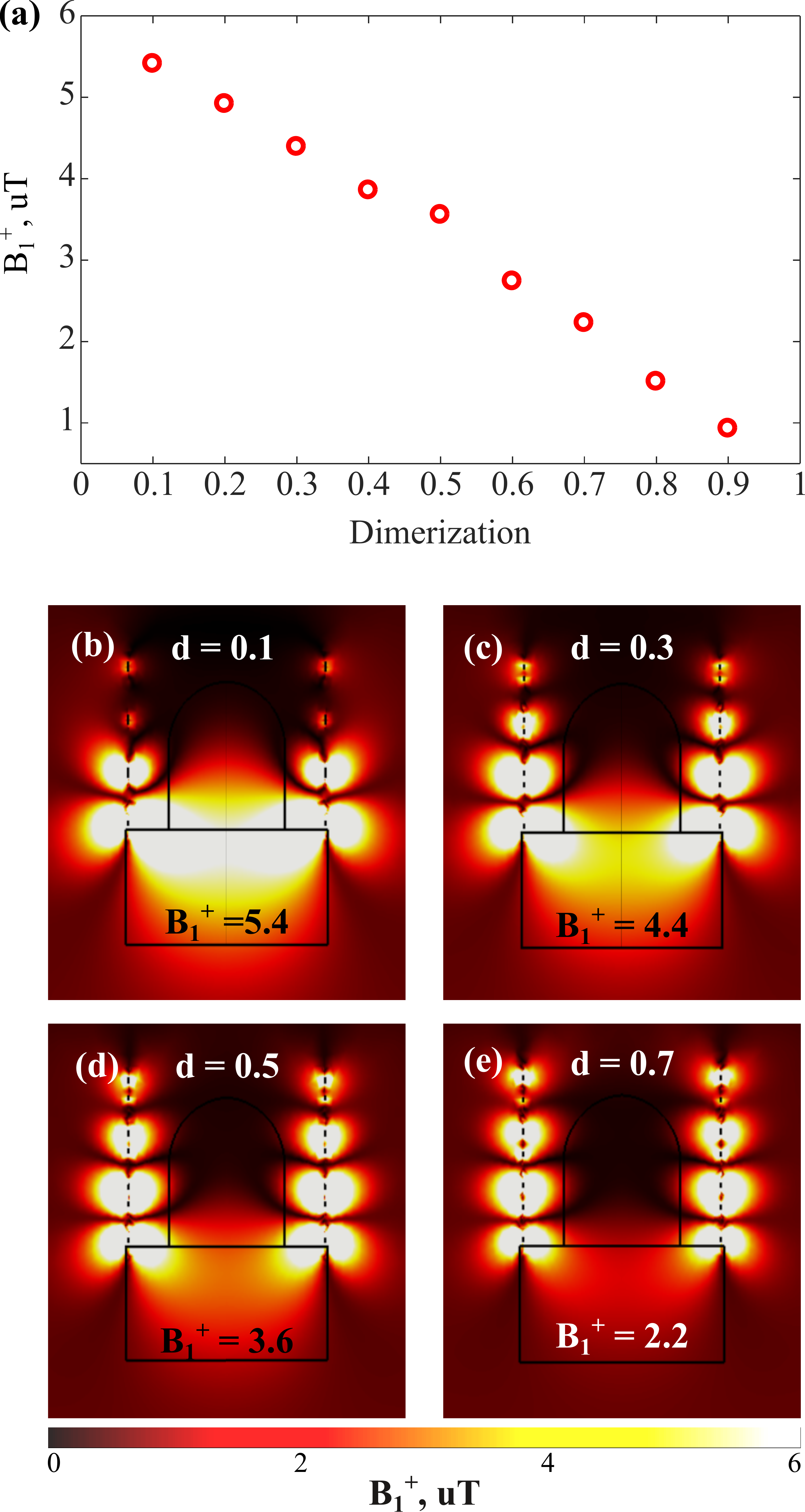}
    \caption{(a) Numerically obtained $B_{1}^{+}$ amplitude dependence on the dimerization degree. (b-r) $B_{1}^{+}$ distributions for different dimerization degrees $d$: (b) $d=0.1$, (c) $d=0.3$, (d) $d=0.5$, and (e) $d=0.7$.}
    \label{fig:bp}
\end{figure*}

In addition to increasing the SNR, the advantage of using the proposed resonator is the transmit efficiency enhancement. This parameter is defined as the magnitude of the clockwise component of the transverse RF magnetic field ($B_{1}^{+}$) normalized by the accepted power. In other words, an increase in transmit efficiency shows how much it is possible to lower the voltage supplied to the scanner to obtain the same signal level. Less input voltage (or power level) also leads to an improvement of the specific absorption rate (SAR), which is a critical factor in estimating the RF safety of the MR procedure. Therefore, we conduct a numerical study of the dependence of the $B_{1}^{+}$ field amplitude on dimerization. The amplitude was calculated in the part of the phantom mimicking the body. As a result, it is seen that lower dimerization ratio values correspond to higher transmit efficiency. The obtained maps [Figs.\ref{fig:bp}(b)-(e)] demonstrate that with increasing dimerization ratio, the localization of the magnetic field $B_{1}^{+}$ near the edge element decreases.

